\documentclass[aps,prl,reprint,amsmath,amssymbols,superscriptaddress]{revtex4-1}

\RequirePackage[mathlines]{lineno} 
\usepackage{graphicx}
\usepackage{bm}
\usepackage{color}
\usepackage{amsmath}
\usepackage{amsfonts}
\usepackage{amssymb}
\usepackage[normalem]{ulem}
\usepackage{verbatim} 

\definecolor{darkGreen}{rgb}{0,0.5,0}
\definecolor{darkBlue}{rgb}{0,0,0.5}
\definecolor{darkRed}{rgb}{0.5,0,0}

\begin{document}

\title{Yield precursor dislocation avalanches in small crystals: the irreversibility
transition}

\author{Xiaoyue Ni}
\email{xni@northwestern.edu}
\affiliation{Division of Engineering and Applied Sciences, California Institute of
Technology, Pasadena, CA, USA}
\author{Haolu Zhang}
\affiliation{Division of Engineering and Applied Sciences, California Institute of
Technology, Pasadena, CA, USA}
\author{Danilo B. Liarte}
\affiliation{Laboratory of Atomic and Solid State Physics, Cornell University, Ithaca,
NY, USA}
\author{Louis W. McFaul}
\affiliation{Physics Department, University of Illinois at Urbana-Champaign, Urbana,
IL, USA}
\author{Karin A. Dahmen}
\affiliation{Physics Department, University of Illinois at Urbana-Champaign, Urbana,
IL, USA}
\author{James P. Sethna}
\affiliation{Laboratory of Atomic and Solid State Physics, Cornell University, Ithaca,
NY, USA}
\author{Julia R. Greer}
\affiliation{Division of Engineering and Applied Sciences, California Institute of
Technology, Pasadena, CA, USA}

\date{\today}
	
\begin{abstract}
The transition from elastic to plastic deformation in crystalline metals shares history
dependence and scale-invariant avalanche signature with other non-equilibrium
systems under external loading such as colloidal suspensions. 
These other systems exhibit transitions with clear analogies to work hardening and
yield stress, with many typically undergoing purely elastic behavior only after
`training' through repeated cyclic loading; studies in these other systems show a
power-law scaling of the hysteresis loop extent and of the training time as the peak
load approaches a so-called reversible-to-irreversible transition (RIT).
We discover here that deformation of small crystals shares these key characteristics:
yielding and hysteresis in uniaxial compression experiments of single-crystalline Cu
nano- and micro-pillars decay under repeated cyclic loading.
The amplitude and decay time of the yield precursor avalanches diverge as the peak
stress approaches failure stress for each pillar, with a power-law scaling virtually
equivalent to RITs in other nonequilibrium systems.
\end{abstract}
	
\maketitle

The mechanical deformation of macroscopic metals is usually characterized by the yield stress, below which the metal responds elastically, and beyond which plastic deformation is mediated by complex dislocation motion and interactions. In small-scale crystals, dislocation activities manifest as avalanches, with characteristic discrete strain bursts in the stress-strain response of the sample~\cite{Uchic:2004, Greer:2005, Dimiduk:2006}. The avalanches exhibit complex scale invariant behavior on wide length scales and time scales~\cite{Miguel:2001, Dimiduk:2006}. The yield stress depends on the history of the sample: if the sample were unloaded and then reloaded during plastic flow, the previous maximum stress would become the current yield stress, below which there are no deviations from linear-elastic response, with the flow and yield stresses always increasing, {\it i.e.} work hardening~\cite{Lubliner:2008}. The elastic-to-plastic transition in crystals finds theoretical analogies to many non-equilibrium material systems~\cite{Sethna:2017}: dilute colloidal suspensions~\cite{Pine:2005, Corte:2008}, plastically-deformed amorphous solids~\cite{Regev:2015, Leishangthem:2017, Jeanneret:2014, Nagamanasa:2014}, granular materials~\cite{Rogers:2014, Mobius:2014, Slotterback:2012}, and dislocation-based simulations of crystals~\cite{Zhou:2014}. In all these other systems, the loading-unloading hysteresis disappears only after repeated cycling to the maximum stress, coined as material training. These systems exhibit power laws and scaling in the limit that the maximum stress approaches a critical value, the so-called reversible-irreversible transition (RIT), which separates trainable and untrainable regimes. For crystals, the nonelastic reloading behavior is in reminiscence of fatigue, in which plastic training is characterized by cyclic strain hardening, an evolution of hysteresis loops, and the emergence of well-defined dislocation microstructures~\cite{Grosskreutz:1971, Basinski:1992}. However, the immediate elastic-nonelastic asymmetry in the unloading-reloading process lies in the realm of abnormal fatigue behavior, such as the anomalous Bauschinger effect, which has only been observed before in polycrystalline metals~\cite{Cleveland:2002}, small system sizes with unconventional microstructures~\cite{Xiang:2006, Brown:2010, Bernal:2015, Rajagopalan:2007} or presence of strong strain gradients~\cite{Liu:2013}, or single crystals during the initial elastic loading~\cite{Ni:2017}. In this work, we discover that sub-micron- and micron-sized metals display the same RIT, with the training hysteresis reduced in larger sample volumes. We begin by showing that the `textbook description' of yield stress and work hardening is fundamentally violated even for metallic single-crystalline micro- and nano-pillars under uniaxial plastic deformation.

\begin{figure}[!ht]
	\centering
	\includegraphics[width=\linewidth]{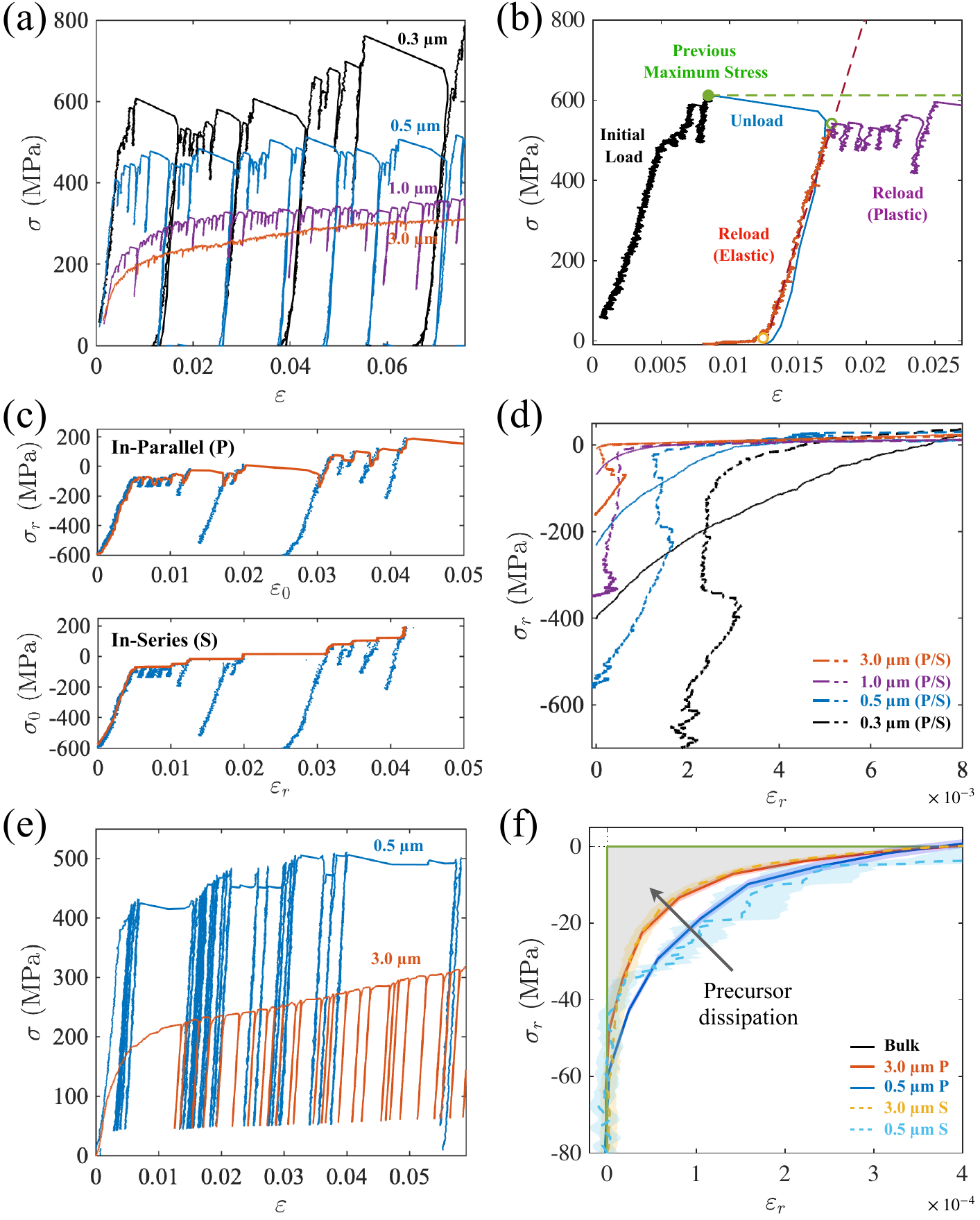}
	\caption{Precursor avalanches present in the uniaxial quasistatic and
	unload-reload cyclic compression experiments on single crystalline Cu pillars.
	(a) Representative stress-strain data for displacement-controlled (DC)
	compression experiments on different diameter pillars.
	(b) A close-up of the first fast-avalanche induced unloading-reloading process in the
	300 nm diameter pillar DC compression test. The first reloading starts to deviate from linear elastic response at a strain of $\sim0.017$,
	while at a stress lower than the updated `yield stress', defined as the previous
	maximum stress.
	(c) The in-parallel (P) stress ($\sigma_r$) and in-series (S) strain ($\varepsilon_r)$ reconstruction of the reloading process marked in (b), where dots and solid lines represent raw and interpolated data separately.
	(d) The non-Hookean reconstruction results obtained from displacement-controlled tests on seven identically-prepared pillars for each size of micropillars.
	(e) Sample stress-strain and (f) the reconstructed non-Hookean stress-strain for
	two representative load-controlled (LC) unload-reload compression experiments on 3 $\mu$m and 500
	nm diameter pillars.
	The area of the shaded region represents the precursor dissipation for 3 $\mu$m
	pillars.}
	\label{fig:precursors}
\end{figure}

Fig.~\ref{fig:precursors}(a) shows typical true stress-strain responses of
displacement-controlled (DC) compression of single-crystalline
$\langle 111 \rangle$-oriented Cu nanopillars with diameters of 300 nm, 500 nm,
700 nm, 1 $\mu$m, and 3 $\mu$m.
This plot reveals multiple discrete strain bursts, which have been shown to
correspond to dislocation avalanches that are triggered from depinning events
during plastic flow~\cite{Zaiser:2006}.
Some occasional strain bursts are also present during the post-avalanche reloading
processes at stresses lower than the current `yield stress', which is defined as the
previous maximum stress that triggered the most-recent avalanche unloading event,
exemplified in Fig.~\ref{fig:precursors}(b) for the 300 nm diameter pillar test.
The presence of such pre-yield avalanches contrasts with the conventional definition
of history-dependent yield point in metals that strictly separates the purely elastic
behavior upon unloading and reloading from irreversible plasticity.
The plastic strain that occurs below the previous maximum stress is the yield-precursor
strain.

As the occurrence of avalanches upon reloading is stochastic in small-scale crystals, we apply two types of stress-strain reconstruction to average all the reloading curves as a measure of the ensemble precursor deviation from the `peak stress' yielding. Fig.~\ref{fig:precursors}(c) demonstrates the in-series and in-parallel reconstruction using the reloading process marked in Fig.~\ref{fig:precursors}(b). We first shift the origin of each reloading process such that the stress is zeroed at the previous maximum stress and the strain is zeroed at the beginning. We interpolate and average the reloading response $\sigma_r$ (in-parallel) or stress $\varepsilon_r$ (in-series), along the monotonically increasing strain $\varepsilon_0$ (in-parallel) or stress $\sigma_0$ (in-series). 
Fig.~\ref{fig:precursors}(d) shows the reconstruction results obtained from displacement-controlled tests on seven identically-prepared pillars for each size of micropillars. We have subtracted the elastic strain to emphasize the plastic precursor behavior (See SI Sec. S2 for details of the reconstruction procedure \footnote{See supplemental material [url] for additional information, which includes refs.~\cite{Maass:2015, Sneddon:1965, Uchic:2009, Cui:2015, Jennings:2011, Agnihotri:2015, Nicola:2006, Marquardt:1963, Press:1986}}).

In the experiments presented here, larger precursor strain is prevalently observed in smaller pillars. However, we observe that the larger pillars that are
monotonically loaded under displacement control generally produce shorter avalanche
strains~\cite{Csikor:2007, Brinckmann:2008} and are less frequently spontaneously unloaded by the instrument
compared with the smaller pillars. The emergent effect
of system size on precursor avalanche behavior, where `system size' refers to the
overall pillar volume, might arise from the variable unloading conditions.
We conduct load-controlled (LC) compression experiments with several prescribed
unload-reload cycles interrupting the quasi-static compression to investigate the size effect.
The maximum stress increases 5 MPa per cycle, which equals to a quasistatic ramping
rate of $\sim$ 1.4 MPa$/$s.
Fig.~\ref{fig:precursors}(e) shows such unload-reload stress-strain response of
representative 500 nm and 3 $\mu$m diameter Cu pillars, and
Fig.~\ref{fig:precursors}(f) compares their reconstructed yield-precursor stress-strain response.
The types of precursor avalanches that we observe during the deformation of small
micropillars that extend over $\sim 10^{-4}$ strains at precursor stresses that are
$\sim 5\%$ (20 MPa) lower than the previous maximum stress ($\sim 400$ MPa)
would pose significant corrections to Hookean elastic behavior if they persisted to
macroscopic systems.

We numerically evaluate the energy dissipation per volume reduced by precursor
avalanches in comparison with the conventional plastic behavior, the precursor
dissipation, from an integral over the reconstructed stress-strain hysteresis,
$U = - \int \sigma_r d \epsilon_0$, indicated by the shaded area in
Fig.~\ref{fig:precursors}(f) for 3 $\mu$m diameter samples.
We observe larger precursor dissipation of $\sim 60$ kPa in the smaller 500 nm
diameter pillars than $\sim 4$ kPa in the larger 3 $\mu$m diameter samples, which
suggests that the precursor avalanches may disappear in macroscopic samples.
This is different from finite-size effect in statistically averaged distributions, where
individual avalanches are hard to resolve in bulk or in high-symmetry
crystals~\cite{Weiss:2015}.
Since we measure the ensemble hysteresis, which is in nature a sum of the
dissipation, small avalanches below the resolution of the instrument will still be
properly incorporated.
Perhaps this explains why precursor avalanches have not been thoroughly
examined in existing literature.

\begin{figure}[!ht]
	\centering
	\includegraphics[width=\linewidth]{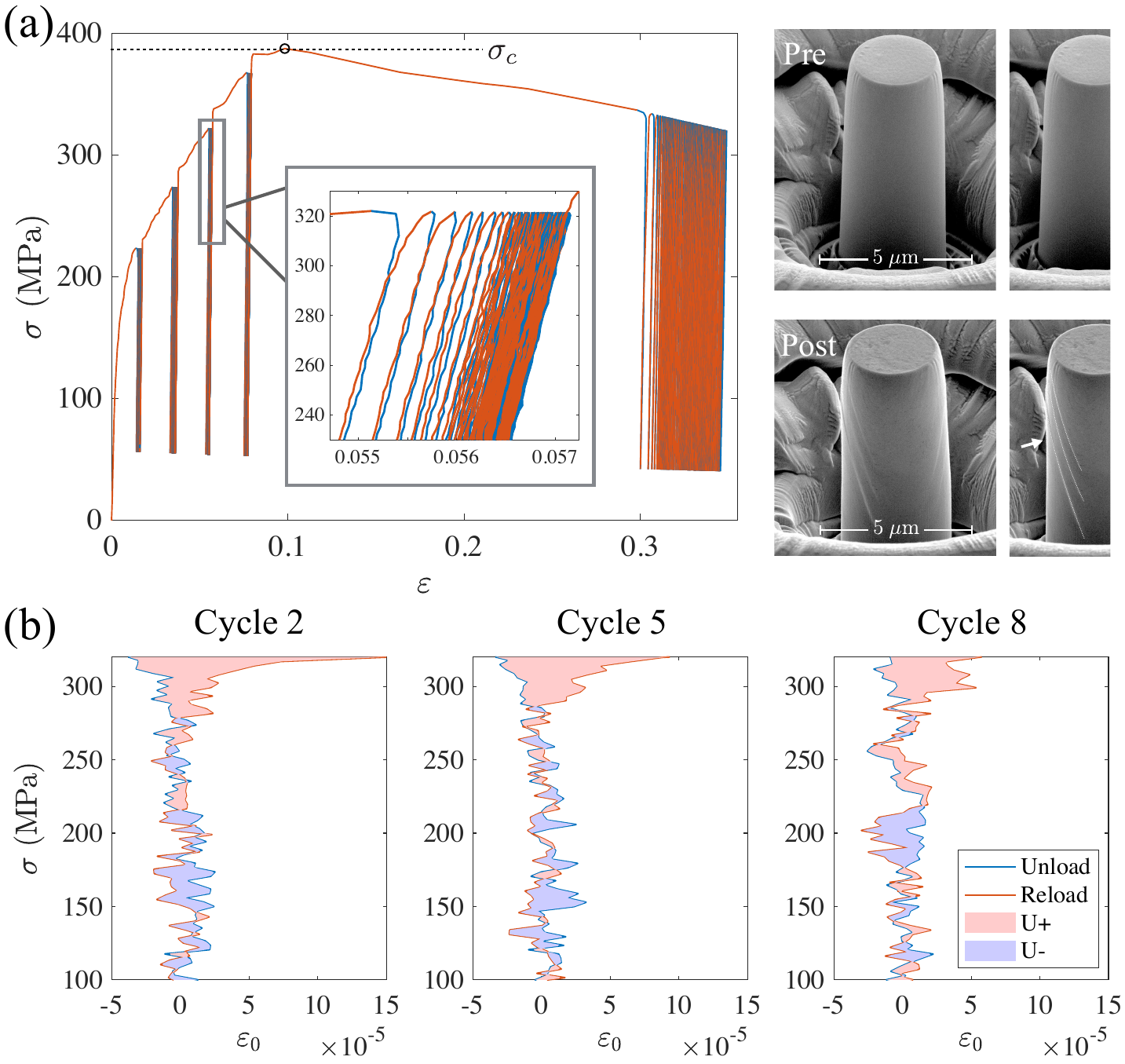}
	\caption{Precursor avalanches trained over cyclic loading in micro-pillars.
	(a) Left: Estimated true stress-strain response from a LC training experiment on a 3
	$\mu$m-diameter Cu pillar.
	Unloading and reloading stress-strain curves are marked in blue and red,
	respectively.
	The maximum stress is increased in five steps; the fifth step reaches the failure
	stress $\sigma_c$.
	At each step, 100 unload-reload cycles are prescribed.
	Right: pre- and post-test scanning electron microscope (SEM) images of this
	sample with the arrow and dashed line denoting the crystallographic slip lines
	on parallel planes characteristic of dislocation avalanches and glide in the
	zoom-in images on the side.
	(b) The drift-corrected stress vs. strain (See SI Sec. S4 for details~\cite{Note1}) during the
	2nd, 5th, and 8th cycles from data shown in (a) loaded to a maximum of
	$\sim 340$ MPa.
	Shaded area represents the energy dissipated through precursor avalanches,
	which decreases over cyclic loading.}
	\label{fig:precursors_training}
\end{figure}

We conduct LC cyclic training experiments to study how the precursor hysteresis
changes under repeated loading to the same maximum stress, analogous to
experiments on other non-equilibrium systems~\cite{Pine:2005, Corte:2008}.
We choose 3 $\mu$m diameter single crystalline Cu pillars as the primary
experimental system because it is sufficiently large amongst the `small-scale'
counterparts to exhibit failure under quasistatic loading as well as relatively
deterministic precursor avalanche behavior.
Fig.~\ref{fig:precursors_training}(a) shows the estimated true stress-strain data from one
representative training experiment on the left along with the scanning electron
microscope (SEM) images of a typical pillar pre- and post-compression on the
right.
The failure stress, $\sigma_c$, or the stress beyond which the
samples are no longer able to support additional applied load, is defined as the global maximum stress at $\sim 390$ MPa.
Above this stress, the sample continually deforms plastically at a constant stress%
~\footnote{%
The perfect failure mode is not observed in the LC compression experiments
because a dynamic increase in engineering stress is required to compensate for
the plastic deformation induced contact area change of the pillar loaded in
$\left< 111 \right>$ high-symmetry direction to maintain the true applied stress
above the critical stress level.}.
In the representative experiment, we prescribe five cyclic stress steps
with maximum engineering stress from 228 MPa ($\sim 0.57 \sigma_c$) to 452 MPa ($\sim 1.15 \sigma_c$) at equal intervals of 56 MPa ($\sim 0.14 \sigma_c$).
In each stress step, we apply 100 unload-reload cycles, during which the sample
is loaded to the same maximum stress and unloaded to a minimum of 56 MPa to
maintain contact between the compression tip and the sample.
We investigate the yield precursor dissipation evolution over all cycles at each
stress step.
Fig.~\ref{fig:precursors_training}(b) shows the 2nd, 5th and 8th cycles of
drift-corrected data (See SI Sec. S4 for details~\cite{Note1}) cycled to $\sim 320$ MPa in
Fig.~\ref{fig:precursors_training}(a), with precursor dissipation indicated by the
shaded areas.

\begin{figure}[!ht]
	\centering
	\includegraphics[width=\linewidth]{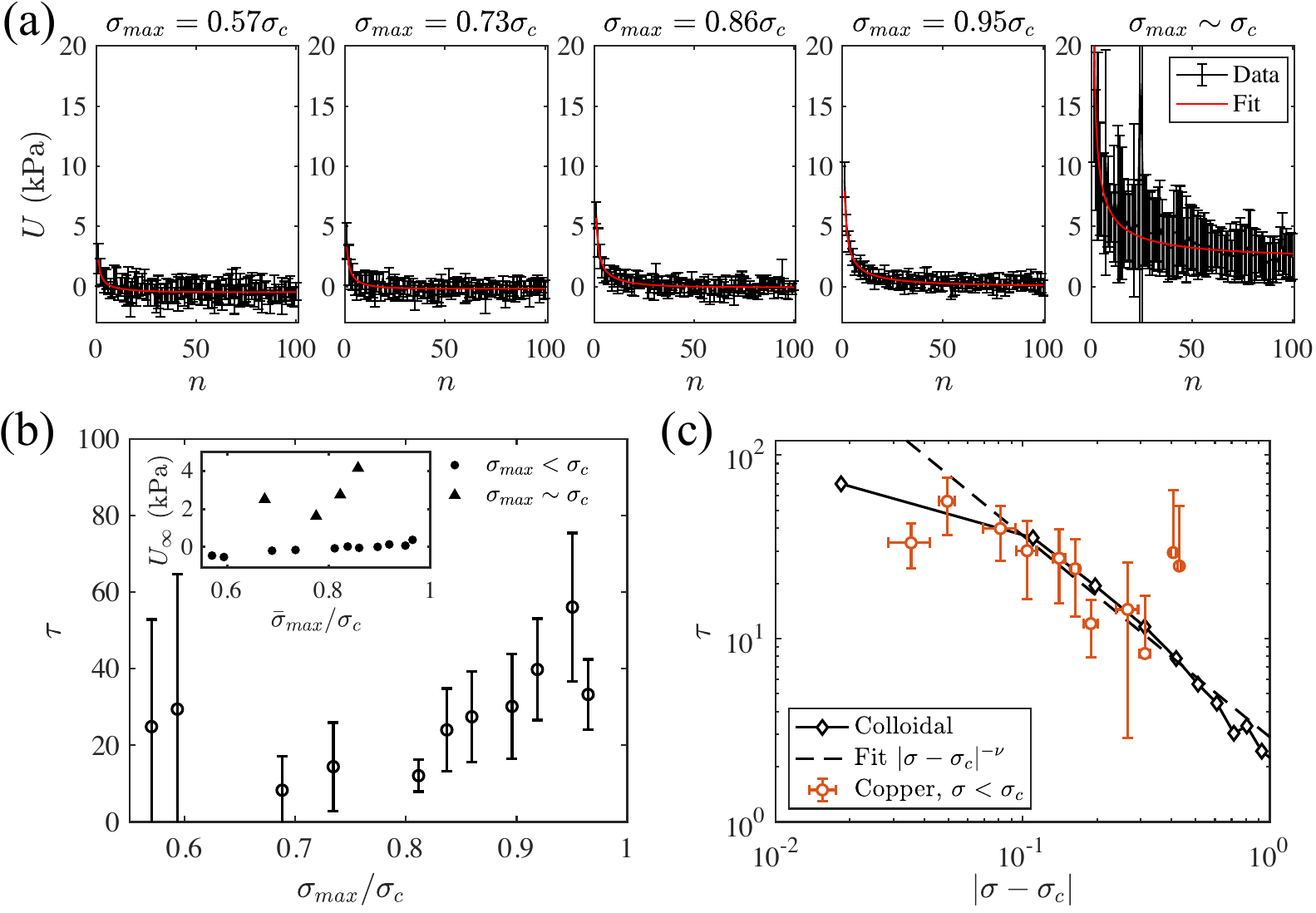}
	\caption{Training experimental results showing precursor dissipation activity
	at different maximum stresses.
	(a) The precursor dissipation energy $U$ at each representative maximum
	stress that shows its decay with the number of loading cycles, $n$.
	(b) Characteristic decay time $\tau$ versus the normalized maximum stress
	$\sigma_\text{max}$ estimated for 3 $\mu$m diameter copper pillars.
	Inset shows that the estimated steady-state $U_\infty$ is close to zero below
	the critical stress $\sigma_c$ and abruptly increases to $\sim 2$ -- 4 kPa when
	$\sigma_\text{max}$ reaches $\sigma_c$.
	The deviation of the final point in (b) and the corresponding points in
	(c) from the expected power-law divergence is probably due to our method
	for estimating the steady-state value (See SI Sec. S7~\cite{Note1}).
	(c) A direct comparison of dislocation RIT behavior gleaned from the Cu
	micropillar compression experiments with that reported for a colloidal particle
	system in a sheared suspension~\cite{Corte:2008}, which provides evidence for a
	divergence of necessary cycle time $\tau$ to reach a reversible state, close
	to the critical failure stress $\sigma_c$.}
	\label{fig:training}
\end{figure}

We apply the multistep cyclic load function spanning the stress range 0.5 -- 1.0
$\sigma_c$ to twenty-four identically prepared samples.
It is reasonable to assume that for a cycle at a specific maximum stress, the
intrinsic precursor dissipation behavior is equivalent across all samples within
statistical variation.
Fig.~\ref{fig:training}(a) shows the average and standard error of the precursor
dissipation as a function of cycle number for increasing maximum stress.
These plots unambiguously demonstrate the training phenomenon: the precursor
hysteresis decays with cycling. Increasing the maximum stress triggers new
precursor avalanches and new training cycles.
Below the catastrophic failure stress $\sigma_c$, the precursor dissipation
virtually vanishes.
Post the failure stress, the hysteretic dissipation continues beyond the
prescribed 100 stress cycles, which indicates that the training is incomplete.

We characterize the decay of precursor dissipation, $U$, versus number of cycles,
$n$, using a fitting function $U_f (n)$~\cite{Corte:2008},
\begin{equation}
	U_f (n)
		= (U_0 - U_\infty) \, e^{-n/\tau} n^{-\delta} + U_\infty,
	\label{eq:Ufunction}
\end{equation}
where $U_\infty = U_f(n \rightarrow \infty)$ is the estimated steady-state dissipation.
$U_0$ is the initial dissipation.
The power-law decay of $U_f$ hints at the fluctuation behavior near the critical point.
This analysis reveals that the catastrophic failure stress $\sigma_c$ in these
experiments can be associated with the reversible-to-irreversible transition (RIT)
critical stress.
This association is corroborated by the non-zero limiting dissipation $U_\infty$ for a
maximum stress amplitude of $\sigma_\text{max} \geq \sigma_c$.
We approximate the long-term decay at the step at
$\sigma_\text{max} \sim \sigma_c$ as critical behavior and fit the precursor
dissipation $U(n)$ using the simple power-law function, $U_f^\prime (n) =
U_f (n; \tau \rightarrow \infty, U_\infty \rightarrow 0) = U_0 n^{-\delta}$, and estimate
the exponent $\delta$ be 0.68.
A separate fit for $\delta$ at different maximum stress gives an average exponent
with standard deviation fluctuation $\delta = 0.70 \pm 0.18$.
We apply the fitted power-law exponent $\delta = 0.70$ to determine $\tau$ for the
remaining stress steps.
Additional fitting details are provided in SI Sec. S5.

We find, unlike granular systems~\cite{Garcia:2005} but like the colloidal systems, that the
dislocation avalanches mostly disappear during the unloading branch (and hence,
at the reversibility transition, also on the loading branch).
This observed behavior could simply reflect a typical dislocation pinning stress
large compared to the failure stress.
Modifying the mean-field model which studies hysteresis in a granular
system~\cite{Bandi:2017}, we incorporate the exponential decay rate $\tau$, and predict
$\delta=1$ (See SI Sec. S6 for details~\cite{Note1}).
The theoretical exponent, however, is far outside our statistical errors for the
collective fit, but within the fluctuations for $\delta$ fit separately for different
$\sigma_\text{max}$.

Fig.~\ref{fig:training}(b) shows that the decay time constant of precursor
hysteresis $\tau$ increases with maximum stress $\sigma_\text{max}$.
Inset shows that the estimated steady-state $U_\infty$ is close to zero below the
critical stress $\sigma_c$ and abruptly increases to $\sim 2$ -- 4 kPa when $\sigma_{max}$
reaches $\sigma_c$~\footnote{The true stress estimates are not reliable post the plastic collapse, as the true stress-strain conversion (See SI Sec. S1~\cite{Note1}) holds only in the conditions that (i) the lattice orientation does not change significantly and (ii) deformation remains approximately homogeneous on the pillar scale.}. 
Plotting the characteristic time scale, $\tau$, as a function of proximity to critical
point on a log-log scale in Fig.~\ref{fig:training}(c), we find a striking resemblance
to the colloidal suspension systems, which indicates that stress-driven dislocations
in small-scale metals exhibit RIT behavior similar to that seen in sheared colloidal
particles~\cite{Corte:2008}. Note that $\tau$, which is related to the number of cycles in our experiments, is not a true timescale of the system.
The deviation of the final point in Fig.~\ref{fig:training}(b) and (c) from the
expected power-law divergence is probably due to our method for estimating the
steady-state value (See SI Sec. S7~\cite{Note1}).

Analogous to the colloidal suspension systems, it is plausible that at low stresses,
the strongly interacting dislocations in the pillars may rearrange themselves into a
stable configuration as the system reloads the first time.
At higher peak stresses, the dislocation rearrangements in one cycle may trigger a
cascade of further avalanches in subsequent cycles.
In small-scale crystalline plasticity, the RIT corresponds to the stress at which
additional cycling continues to plastically deform the system with no additional
applied forces, which corresponds to the failure stress.
We can speculate about the relation between the critical behavior of the precursor
avalanches observed here and the power-law distribution of dislocation avalanches
observed in nano- and micropillars under monotonic loading.
The precursor avalanches at an RIT usually diverge in size only near the failure
stress.
Plasticity avalanches under monotonic loading are debated to be associated with a `stress-tuned criticality' \cite{Friedman:2012, Dahmen:2009} or a jamming transition ~\cite{Ispanovity:2014}, either of which exhibit a power-law scaling with a cutoff in the
avalanche size distribution that diverges only as the stress approaches the `failure
stress' --- precisely as one would expect for the approach to an RIT.

In this work, we bring attention to the overlooked signature of yield precursor
avalanches in nanomechanical experiments.
We show that the amount of dissipation due to yield precursor avalanches decays
over repeated stress training cycles. 
We find that the characteristic decay time increases with the applied maximum
stress.
The apparent divergence of the time constant at a maximum stress near the
quasistatic failure stress indicates that the flow transition of the dislocation system
is fundamentally an RIT.
This is the first time that this effect has been shown in any crystalline material
experimentally.
Prior studies have only focused on amorphous materials and attributed RIT
behavior to many disordered or short-range ordered material systems.
Our work extends the universality of RIT to include crystals.
The training and RIT behavior has potential connections with cyclic fatigue and the transition from rapid hardening to saturation hardening at bulk scales, e.g. shakedown and ratcheting~\cite{Suresh:1998}, wherein the dislocations microstructures evolve from mutual trapped bundles into distant loop patches~\cite{Grosskreutz:1971, Basinski:1992}. However, we demonstrate that size effect is not negligible, which corroborates with the lack of prior research on training effects and precursor avalanches at large scales.
Nanomechanical experiments have been intensively explored as a powerful
methodology to study the fundaments of crystal deformation, but the understanding
of the dislocation plasticity in terms of RIT was hitherto lacking because people had
only focused on quasi-static experiments which are not efficient in resolving
history-dependent dissipative features of materials.
Our work may inspire novel approaches to study plasticity, fatigue, and catastrophic failure
in crystalline materials governed by complex dislocation dynamics.

\begin{acknowledgments}
J.R.G. and X.N. acknowledges financial support from the U.S. Department of
Energy's Office of Basic Energy Sciences through grant DESC0016945.
J.P.S. and D.B.L. acknowledge the financial support of the U.S. Department of
Energy's Office of Basic Energy Sciences through Grant DE-SC0006599 and NSF
grant DMR-1719490.
K.A.D acknowledges NSF grant CBET 1336634.
We thank Stefano Zapperi, Giulio Costantini, D. Zeb Rocklin, Archishman Raju,
and Lorien Hayden for helpful discussions.	
\end{acknowledgments}

\bibliography{Crystal_RIT}
	
\end{document}